# QoE Modeling Associated with QoS Impairment Parameters in 5G Networks Using AHP Decision Making Technique

Therdpong Daengsi
*Sustainable Industrial Management Engineering*
*RMUTP*
Bangkok, Thailand
therdpong.d@rmutp.ac.th

Patsita Sirawongphatsara
*Information Technology*
*KMUTNB*
Bangkok, Thailand
patsita_si@rmutto.ac.th

Phisit Pornpongtechavanich
*Information and Communication Technology for Education*
*KMUTNB*
Bangkok, Thailand
phisit.kha@rmutr.ac.th

*Abstract*—This paper aims to propose the Quality of Experience (QoE) models based on the expectation and/or the perception of 5G users to evaluate for Mean Opinion Score (MOS) for real-time or interactive services/applications with high reliability. Therefore, based on the fundamental QoE concept, the Analytic Hierarchy Process (AHP) decision making technique has been applied. Seven participants or evaluators were recruited for the interview remotely via phone calls. From the analysis using AHP technique, the coefficients or weights for the impairment parameters, consisting of loss, delay and jitter were obtained. Also, the weights of the focused parameters reveal that according to the participants' opinions, loss can generate a worse quality issue when compared with delay and jitter.

## I. INTRODUCTION

### A. Background and Significance

5G networks are becoming the major mobile communication networks globally in the near future. Although, 5G has a better performance than previous mobile networks obviously, packet loss, packet delay, and jitter are still found. These are the impairment parameters that usually impact on the quality of experience (QoE) of users who use real-time services or applications, (e.g., video calls), see Fig. 1 [1]. When the QoE level is usually based on a user's satisfaction, which is related to the users' expectation as well.

The network parameters as in Fig. 1 are also known as the quality of service (QoS) parameters. They can be applied for QoE calculation using the conceptual function as in (1)-(2), which is adopted from [2]. However, functions for QoE calculation referring to QoS parameters were presented widely, such as [3-4], where $QoE_{Overall}$ is the overall QoE score, $QoE_{Loss}$, $QoE_{Delay}$ and $QoE_{Jitter}$ are the Mean Opinion Scores (MOS) associated with loss, delay and jitter respectively, and *a, b* and *c* are the coefficients for each impairment factor. However, due to the brief literature review based on this concept, the overall QoE model referring to the impairment parameters in 5G networks and the expectation of many Thai users, particularly by using the Analytic Hierarchy Process (AHP) technique, has not been studied yet. Therefore, this is the rationale for conducting this study to propose a mathematical QoE model based on the aforementioned criteria.

$$QoE_{Overall} = f(Loss, Delay, Jitter) \qquad (1)$$

$$QoE_{Overall} = a \times QoE_{Loss} + b \times QoE_{Delay} + c \times QoE_{Jitter} \qquad (2)$$

### B. Overview on 5G

5G is the fifth generation of mobile communications that is also called the International Mobile Telecommunications – 2020 (IMT-2020) recommended by the International Telecommunication Union (ITU) [5-6]. 5G has been designed for much higher performance as shown in Fig. 2 [7], in order to support the main use cases that consists of Enhanced Mobile Broadband (eMBB), Ultra-Reliable and Low Latency Communications (uRLLC) and Massive Machine-Type Communications (mMTC) as shown in Fig. 3 [8].

### C. QoS Impairment Parameters: Loss, Delay and Jitter

As mentioned in [9], loss, delay and jitter are the classical QoS parameters. These are the impairment factors that usually impact the quality provided by real-time or multimedia services and/or applications. They are like the three members of evils in IP networks that can be described as follows:

- Loss normally occurs when packets are transmitted from a source but some of them are dropped before reaching the destination [10]. It may occur from buffer overflow due to network congestion does happen [11].
- Delay is the difference between times taken by the transmitted packets from source to destination, including propagation and transmission delay, for example [12]. Delay may not impact non-interactive services/applications but it usually impacts interactive services and applications.
- Jitter is the delay time variation of sequent packets transmitted in a network [13]. This can impact both non-interactive and interactive services and applications. A small amount of jitter may be mitigated by using jitter buffer [14].

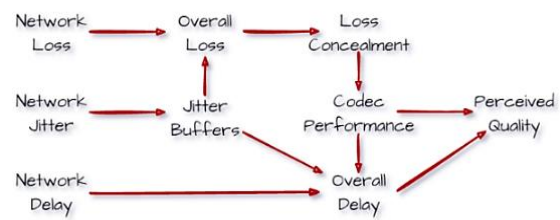

Fig. 1. Major impairment factors referring to perceived quality of users [1]

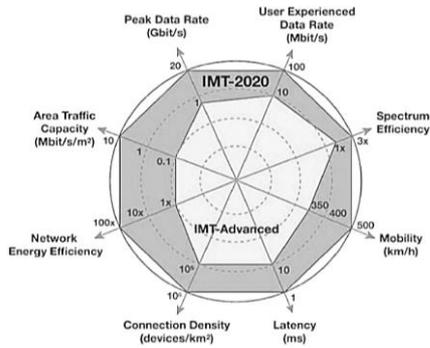

Fig. 2. 5G spider chart [7]

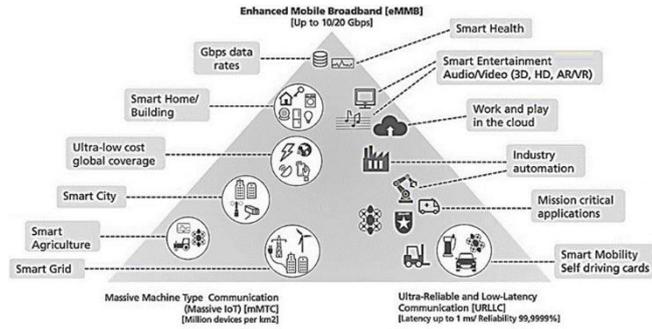

Fig. 3. 5G spider chart [8]

*D. AHP*

Several methods based on Multiple Criteria Decision Making (MCDM) or Multi-Attribute Decision Making (MADM) have been utilized to facilitate cases in which at least two criterions are considered and evaluated in the decision-making process. From those methods, the AHP is an easier decision-making technique to be used to evaluate the related criteria involved with the decision [15]. Pair-wise comparisons referring to criterions are performed based on 9-level scale in AHP process before generating normalized weights to each parameter by capturing experts' knowledge of phenomena under consideration [16-17]. As mentioned in [18], AHP is very often applied for deriving weights associated with network parameters.

*E. Related Works*

For QoE metrics, the Mean Opinion Score (MOS) is one of the most acceptable metrics [19]. MOS is traditionally calculated from the subjective evaluation results using the 5-point scale (where 5 = excellent, 4 = good, …, and 1 = bad). This QoE approach is called a subjective method. Then the objective measurement methods have been developed based on mathematical models (e.g., Simplified E-model and E-model) [20]. From the brief survey in this study, many QoE models associated with QoS parameters were proposed [3-4]. Some of the interesting models can be presented as in (3) [21]. This model is for voice over IP (VoIP) quality evaluation.

$$R_{Simplified} = R_o - I_{delay} - I_{codec\&loss} - I_{jitter} \quad (3)$$
$$I_{delay} = 0.024 \times delay + 0.11 \times (delay - 177.3) \times H(delay - 177.3) \quad (4)$$
$$I_{codec\&loss} = A + B \times \ln(1 + C \times P/100) \quad (5)$$
$$I_{jitter} = C_1 \times H^2 + C_2 \times H + C_3 + C_4 \times e^{-T/K} \quad (6)$$

Where $R_{Simplified}$ is 93.2, $I_{delay}$ is a factor that is impacted by delay calculated with the Heavyside function (H(x)) by using (4), $I_{codec\&loss}$ is a factor referring to packet loss and codec effects calculated with A, B and C of 11, 40 ms and 10 respectively for G.729 audio codec and the total loss probability (P) by using (5), and $I_{jitter}$ is a factor associated with jitter effects calculated with $C_1$, $C_2$, $C_3$ and $C_4$ of -15.5, 33.5, 4.4 and 13.6 respectively, while H, T and K are 0.6, 40 ms and 30 respectively for G.729 codec, and H is the Pareto distribution factor of 0.55-0.9. However, after obtaining the result of $R_{Simplified}$, it must be converted from 100-point scale into 5-point scale using (7) [22], where R is $R_{Simplified}$.

$$MOS = \begin{cases} 4.5 & ; R>100 \\ 1+0.035 \times R + R(R-60)(100-R)7 \times 10^{-6} & ;0<R<100 \\ 1 & ; R<0 \end{cases} \quad (7)$$

However, the model as in (3) associated with three network impairment parameters only, whereas it did not apply AHP technique. For the other interesting QoE models referring to video quality developed using AHP are shown as in (8)-(9), where ARS in (8) is auto resolution scaling [23].

$$QoE_{Network} = 0.26 \times Loss + 0.55 \times Jitter + 0.07 \times Throughput + 0.12 \times ARS \quad (8)$$
$$QoE_{Application} = 0.26 \times (Bit\ rate) + 0.63 \times (Frame\ rate) + 0.11 \times (Resolution) \quad (9)$$

One can see that (3) is a QoE model for VoIP quality evaluation based on impairment parameters but it is very complex and not related to AHP technique. While (8)-(9) are the QoE models developed by using AHP technique but they are not related to three impairment parameters. Thus, there is a research gap for the development of the new QoE model related to real-time services and applications operated on 5G networks using AHP.

## II. METHODS

According to the AHP technique, this study requires the advice from a group of evaluators or experts to avoid the subjective bias from a single evaluator. However, 5G technology is new for Thai people, it is difficult to find a group of 5G experts. Therefore, the evaluators or participants were recruited from the members of the webpage of telecom community in Thailand and some network engineers, who were interested in 5G technology and realized or were aware about the effects of the QoS impairment parameters, instead. From the recruitment, there were seven evaluators, consisting of six 4G/5G super users and one network engineer who hold a master degree in Information Technology. The average age of them was 31.29±9.30 years.

For the interview session, each one was interviewed via mobile phone to evaluate a series of pairwise comparisons (consisting of loss vs. delay, loss vs. jitter, and delay vs. jitter, see Fig. 4) for identifying how important a specific QoS impairment parameter is within each pair, using a 9-point scale. Then the data from the interview obtained from every interviewee were gathered and analyzed, as presented in Section III.

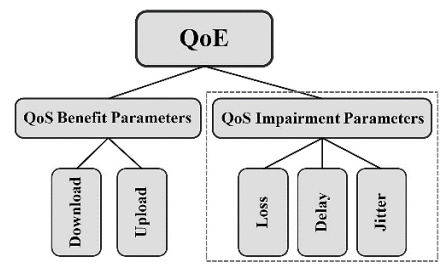

Fig. 4. QoE and its QoS related parameters [5]

## III. RESULTS AND DISCUSSION

After obtaining the data from the group of seven evaluators in total, the data were processed using AHP technique as show in Table I and II. Then the average values as shown in the last column in Table II have been applied as the proposed coefficients into the (2). Then it becomes the proposed QoE model as shown in (10), where $MOS_{Overall}$ is overall MOS value, while $MOS_{Loss}$, $MOS_{Delay}$ and $MOS_{Jitter}$ are MOS values referring to the classic impairment parameters, loss, delay and jitter respectively.

Due to loss, delay and jitter parameters, this QoE model is appropriate to real-time or interactive services and/or applications only. Focusing on each weight for each impairment parameter, one can see that the weight of 0.55 of loss becomes the significant impact factor on the interviewees' opinions, while the weights of 0.25 and 0.20 of delay and jitter respectively are smaller impact than loss.

Although this paper presents a new QoE model obtained from the reasonable AHP approach, it requires model performance evaluation that should be conducted in the next study. Furthermore, it should cover more 5G experts and specialists for more reliability and confidence. Besides, this study focused on the impairment factors only, the future study should be extended to the benefit parameters, download and upload speeds. In addition, this study has not revealed the details about $MOS_{Loss}$, $MOS_{Delay}$ and $MOS_{Jitter}$, as recommended to be conducted in the future.

## IV. CONCLUSION

After performing the AHP decision making technique for obtaining the weights or coefficients associated with loss, delay and jitter that are three major impairment parameters in IP networks, including 5G networks, the QoE model referring to those parameters and 5G networks have been proposed. Furthermore, this study has revealed that loss has the biggest impact of 55%, while delay and jitter have the impacts of 25% and 20% only respectively. However, this QoE model requires additional performance evaluation, which can be studied as future work, while the extended study with benefit parameters should be performed as well.

## ACKNOWLEDGMENTS

Firstly, Gratitude to University of Bahrain for organizing DASA'21 and supporting the cost for registration. Thanks to Rajamangala University of Technology Phra Nakhon for supporting this study. Thanks to all participants who involved with the interview. Finally, thanks to Mrs.Anong Lertrakskun and Ms.Cecilia Mei-Yun Oh for English editing.

TABLE I. COMPARISON OF IMPORTANCE OF QOS PARAMETERS

| Importance | Loss | Delay | Jitter |
|---|---|---|---|
| Loss | 1 | 5.74 | 5.48 |
| Delay | 0.95 | 1 | 2.48 |
| Jitter | 0.67 | 1.95 | 1 |

TABLE II. WEIGHT OF LOSS, DELAY AND JITTER FOR QOE MODEL

| Weight | Loss | Delay | Jitter | Average |
|---|---|---|---|---|
| Loss | 0.38 | 0.66 | 0.61 | 0.55 |
| Delay | 0.36 | 0.11 | 0.28 | 0.25 |
| Jitter | 0.26 | 0.22 | 0.11 | 0.20 |

$MOS_{Overall} = 0.55 \cdot MOS_{Loss} + 0.25 \cdot MOS_{Delay} + 0.20 \cdot MOS_{Jitter}$ (10)

## REFERENCES

[1] ETSI, "Telecommunications and internet protocol harmonization over networks (TIPHON) Release 3; End-to-end Quality of Service in TIPHON systems; Part 7: design guide for elements of a TIPHON connection from an end-to-end speech transmission performance point of view," 2002. https://www.etsi.org/deliver/etsi_tr/101300_101399/10132907/02.01.01_60/tr_10132907v020101p.pdf.

[2] F. Liberal, A. Ferro, H. Koumaras, A. Kourtis, L. Sun, and E. C. Ifeachor, "QoE in multi-service multi-agent networks," *Int J Commun Netw Distrib Syst*, vol. 4(2), 183-206, 2010.

[3] E. Liotou, D. Tsolkas and N. Passas, "A roadmap on QoE metrics and models," *Proc. of ICT 2016*, Thessaloniki, Greece, May 2016, pp. 1-5.

[4] D. Tsolkas, E. Liotou, N. Passas and L. Merakos, "A survey on parametric QoE estimation for popular services," *J. Netw. Comput Appl.*, vol. 77, pp. 1-17, 2017.

[5] ITU-T Recommendation Y.3100, "Terms and definitions for IMT-2020 network," 2017.

[6] ITU-R Recommendation M.2150-0, "Detailed specifications of the terrestrial radio interfaces of International Mobile Telecommunications-2020 (IMT-2020)," 2021.

[7] Moniem-Tech, "How Does 5G Achieve 20 Gbps?" 2018. https://moniem-tech.com/2018/06/29/how-does-5g-achieve-20-gbps/

[8] A. Heuberger, "5G technologies for a range of use cases," 2019. https://www.iis.fraunhofer.de/en/magazin/2019/5g_technologies.html

[9] A. D. Tesfamicael, V. Liu, E. Foo, and B. Caelli, "QoE Estimation Model for a Secure Real-Time Voice Communication System in the Cloud," *Proc. of ACSW 2019*, Sydney, Australia, Jan 2019, pp. 86–89.

[10] IGI-Global, "What is Packet Loss Rate." https://www.igi-global.com/dictionary/packet-loss-rate/21753

[11] J. P. Mettilsha, M. K. Sandhya, and K. Murugan, "RPR: Reliable path routing protocol to mitigate congestion in critical IoT applications" *Wirel Netw*. https://doi.org/10.1007/s11276-021-02805-w

[12] IGI-Global, "What is Packet Delay." https://www.igi-global.com/dictionary/packet-delay/21748

[13] T. Daengsi and P. Wuttidittachotti, "Quality of Service as a Baseline for 5G: A Recent Study of 4G Network Performance in Thailand," *Proc. of COMNETSAT 2020*, Batam, Indonesia, Dec 2020. pp. 395-399

[14] V. Joseph, and B. Chapman, Deploying QoS for Cisco IP and Next Generation Networks, Elsevier. https://doi.org/10.1016/B978-0-12-374461-6.X0001-8

[15] F. Stocker, E. G. Villar, K. D. D. Roglio, and G. Abib, "Dismissal: Important Criteria in Managerial Decision-Making," *RAE-Revista de Administração de Empresas (Journal of Business Management)*, vol. 58(2), pp. 116-129, 2018.

[16] M. Sadik, N. Akkari, and G. Aldabbagh, "QoS/QoE Based Handover Decision in Multi-Tier LTE Networks," *Int J Digit Inform Wirel Commun*, vol. 8(2), pp. 133-138, Jun 2018.

[17] L. D. S. Pacheco, "Mobility and Cloud Management in Wireless Heterogeneous 5G Networks," Dissertation, Institute of Technology, Federal University of Para, 2020.

[18] G. Caso, O. Alay, G. C. Ferrante, L. D. Nardis, M. D. Benedetto, and anna brunstrom, "User-Centric Radio Access Technology Selection: A Survey of Game Theory Models and Multi-Agent Learning Algorithms," *IEEE Access*, vol 9, pp. 84417-84464, 2021.

[19] ITU-T Recommendation P.800, "Methods for subjective determination of transmission quality", 1996.

[20] P. Wuttidittachotti and T. Daengsi, "VoIP-Quality of Experience Modeling: E-model and Simplified E-model Enhancement Using Bias Factor," *Multimed Tools Appl*, vol. 76(6), pp. 8329-8354, 2017

[21] D. Nguyen, H. Nguyen and É. Renault, "Performance evaluation of E-MQS scheduler with Mobility in LTE heterogeneous network," *Proc. of ICC 2017*, Paris France, May 2017, pp. 1-6.

[22] ITU-T Recommendation G.107, "The E-model: a computational model for use in transmission planning," Jun 2015.

[23] D. Pal and C. Vanijja, "A No-Reference Modular Video Quality Prediction Model for H.265/HEVC and VP9 Codecs on a Mobile Device," Adv Multimed, vol. 2017, Article ID 8317590, https://doi.org/10.1155/2017/8317590.